\documentclass[aps,twocolumn,pra,reprint,amsmath,amssymb,floatfix,footinbib,superscriptaddress]{revtex4-1}
\usepackage{float}
\usepackage{amsmath}	
\usepackage{gensymb}
\usepackage{amssymb}
\usepackage{yhmath}
\usepackage{bbold}
\usepackage{hyperref}
\usepackage{tikz}
\usepackage{array}
\hypersetup{colorlinks=true}
\usepackage{upgreek}
\usepackage{graphics}
\usepackage{hyperref}
\usepackage{epsfig}
\usepackage{color}
\usepackage{bm}
\usepackage{graphicx}
\usepackage{indentfirst}
\usepackage{ulem} 
\usepackage{blindtext}
\usepackage[caption=false]{subfig}
\usepackage{natbib}

\usepackage{CJK}        
\usepackage{url}
\usepackage{multirow}
\definecolor{clr_K}{RGB}{128,128,128}
\definecolor{clr_Gamma}{RGB}{127.5,0,127.5}
\definecolor{clr_M1}{RGB}{0,128,0}
\definecolor{clr_M2}{RGB}{255,0,255}
\definecolor{clr_M3}{RGB}{0,128,0}
\definecolor{clr_AFMS1}{RGB}{119,172.5,48}
\definecolor{clr_AFMS2}{RGB}{255,125.5,0}
\definecolor{clr_FMS1}{RGB}{0,255,255}
\begin{document}

\title{Interplay of Kitaev Interaction and Off-diagonal Exchanges: Exotic Phases and Quantum Phase Diagrams}

\author{Qiang Luo}
\email[]{qiangluo@nuaa.edu.cn}
\affiliation{College of Physics, Nanjing University of Aeronautics and Astronautics, Nanjing, 211106, China}
\affiliation{Key Laboratory of Aerospace Information Materials and Physics (NUAA), MIIT, Nanjing, 211106, China}
\author{Jize Zhao}
\affiliation{School of Physical Science and Technology $\&$ Key Laboratory of Quantum Theory and Applications of MoE, Lanzhou University, Lanzhou 730000, China}
\affiliation{Lanzhou Center for Theoretical Physics, Key Laboratory of Theoretical Physics of Gansu Province, Lanzhou University, Lanzhou 730000, China}
\author{Xiaoqun Wang}
\email[]{xiaoqunwang@zju.edu.cn}
\affiliation{School of Physics and Institute for Advanced Study in Physics, Zhejiang University, Hangzhou 310058, China}

\begin{abstract}
Aligning with the everlasting search for quantum spin liquids (QSLs),
identifying the QSL in Kitaev magnets has garnered great research interest during the past decade and remains nevertheless an enormous challenge.
One of the major difficulties lies in that Kitaev QSL is typically fragile against competing interactions like off-diagonal exchanges,
which are ubiquitous in real materials due to spin-orbit coupling and crystal-field effect.
This, in turn, gives rise to many intriguing field-induced novel phases and thermal Hall effect.
In this review, we will focus on the interplay of Kitaev interaction and off-diagonal $\Gamma$ and $\Gamma'$ exchanges from a numerical perspective.
This review discusses some representative exotic phases such as $\Gamma$ spin liquid, nematic ferromagnet, spin-flop phase,
and distinct chiral-spin states with spontaneously time-reversal symmetry breaking.
It also presents quantum phase diagrams of anisotropic Kitaev-$\Gamma$ chains that exhibit kaleidoscopes of both ordered and disordered phases.
\end{abstract}


\maketitle


\textit{1. Introduction}.
Hitherto, experimental hunt for, and theoretical prediction of, exotic states of matter such as quantum spin liquid (QSL)
remain a remarkable challenge at the forefront of modern condensed-matter research \cite{Balents2010Nature,Zhou2017RMP,Wen2019NPJQM,Broholm2020Science}.
Initiated by P. W. Anderson more than a half-century ago \cite{Anderson1973},
the concept of QSL has garnered significant attention
due to it goes beyond Landau's paradigm of characterizing phases by spontaneous symmetry breaking and local order parameter.
The QSL is a highly entangled magnetic state featured by fractionalized excitations and continuous spectrum,
and plays a potential role in topologically protected quantum computing \cite{Wen2017RMP}.
For many years it is realized that low-dimensional quantum magnets with geometrical frustration and/or competing interactions are attractive arenas to probably harbor QSLs,
of which prominent examples are Heisenberg model on Kagome lattice \cite{Liao2017PRL,Zhang2020PRB}
and $J_1$-$J_2$ model on triangular lattice \cite{Hu2019PRL,Gong2019PRB} or square lattice \cite{Wang2018PRL,Liu2022SB}.

Owing to the pioneering work by Kitaev in 2006 \cite{Kitaev2006AP},
introduction of the Kitaev interaction has revolutionized research on hunting for spin-liquid physics.
The spin-$1/2$ Kitaev honeycomb model is demonstrated to be exactly solvable
and its ground state is dubbed Kitaev QSL that possesses fractionalized Majorana fermions \cite{Kitaev2006AP}.
This solvability is vital as it provides an exhaustive understanding of the intricate behaviors exhibited by QSL.
Since then, there is a flourishing study of Kitaev materials that are equipped with non-negligible Kitaev interaction
(for reviews, see Refs. \cite{Hermanns2018ARCMP,Takagi2019NPR,Trebst2022PR,Rousochatzakis2024RoPP}).
In the early stages of exploring Kitaev materials, attention was chiefly focused on $4d/5d$ transition metal compounds \cite{Jackeli2009PRL,Singh2010PRB,Sears2015PRB},
among which $\alpha$-RuCl$_3$ is a star material \cite{Sears2015PRB}.
Accompanied by the deeper understanding of the microscopic mechanism of Kitaev interaction,
there is subsequently a growing interest in cobaltates \cite{Songvilay2020PRB,Lin2021NC,Li2022PRX,Zhong2020SA}
and rare-earth chalcohalides \cite{Ji2021CPL}.
Meanwhile, evidences of Kitaev interaction are also revealed in some Ni/Cr-based van der Waals magnets albeit with higher spins \cite{Stavropoulos2019PRL,Zvereva2015PRB,Shangguan2023NP,Taddei2023PRR,Xu2018npjCM,Xu2020PRL,Stav2021PRR}.
Regrettably, the sought-after QSL has not yet been achieved experimentally since, in real materials,
isotropic Heisenberg interaction and off-diagonal exchanges are not fully quelled
and compete with the Kitaev term, usually resulting in magnetically ordered ground states like zigzag ordering \cite{Maksimov2020PRR}.

Before proceeding further, it is worth noting that there are already many excellent reviews pertaining to the Kitaev physics, both theoretically and experimentally.
In particular, Hermanns \textit{et al.} reviewed the Kitaev honeycomb model from the aspects of fractionalization, dynamic correlations, and its material connections \cite{Hermanns2018ARCMP}.
Takagi \textit{et al.} reviewed the concept and realization of Kitaev QSL in candidate materials \cite{Takagi2019NPR}.
Trebst and Hickey reviewed current status of the exploration of representative Kitaev materials \cite{Trebst2022PR}.
Rousochatzakis \textit{et al.} reviewed the luxuriant emergent phenomena beyond Kitaev physics \cite{Rousochatzakis2024RoPP}.

In this short review, we will focus on the exotic phases and quantum phase diagrams studied by us and our collaborators in the realm of so-called $JK\Gamma\Gamma'$ model.
The chief concerns of exotic phases are the gapless $\Gamma$ spin liquid ($\Gamma$SL) in honeycomb $\Gamma$ model \cite{Luo2021NPJ},
the chiral spin state and nematic ferromagnets in spin-1 Kitaev-$\Gamma$ model \cite{Luo2024PRB},
and the chiral-spin ordering and spin-flop phase in $\Gamma$-$\Gamma'$ model \cite{Luo2022PRR,Luo2022PRB}.
The quantum phase diagrams of spin-$1/2$ and spin-1 bond-alternating Kitaev-$\Gamma$ chains are summarized \cite{Luo2021PRB,Luo2021PRR},
followed by a scrutiny of the Kitaev phase \cite{Luo2023PRB}.
In doing so, we aim to offer insights to aid in the research and advancement of unconventional quantum criticality in frustrated magnets.

\textit{2. Model and Relevant Materials}.
In the transition-metal compounds with edge-sharing octahedra,
interplay of spin-orbit coupling and crystal-field effect conspires to bring about bond-directional Ising-type interactions,
causing the formulation of a large family of Kitaev materials on a honeycomb lattice \cite{Jackeli2009PRL}.
Although these materials have different magnetic ground states and display distinguishing magnetic-field induced phenomena,
a paradigmatic $JK\Gamma\Gamma'$ model is demonstrated to describe them legitimately \cite{Rau2014PRL}. The Hamiltonian takes the form:
\begin{align}\label{EQ:JKGGpHc-Ham}
\mathcal{H} =
    & \sum_{\left<ij\right>\parallel\gamma} \Big[J \mathbf{S}_i \cdot \mathbf{S}_j + K S_i^{\gamma} S_j^{\gamma}
    + \Gamma \big(S_i^{\alpha}S_j^{\beta}+S_i^{\beta}S_j^{\alpha}\big)\Big]   \nonumber \\
    & + \Gamma' \sum_{\left<ij\right>\parallel\gamma}
        \Big[\big(S_i^{\alpha} + S_i^{\beta}\big) S_j^{\gamma} + S_i^{\gamma} \big(S_j^{\alpha} + S_j^{\beta}\big) \Big]    \nonumber \\
    & - \sum_i \mathbf{h} \cdot \mathbf{S}_i,
\end{align}
where $S_i^{\gamma}$~($\gamma$ = $x$, $y$, $z$) is the $\gamma$-component of spin operator at site $i$.
On $z$ bonds $(\alpha, \beta, \gamma) = (x, y, z)$, with cyclic permutation for $x$ and $y$ bonds.
$J$ and $K$ are the isotropic Heisenberg interaction and diagonal Kitaev interaction, respectively,
while $\Gamma$ and $\Gamma'$ are symmetry-allowed off-diagonal exchanges.
The last term in Eq.~\eqref{EQ:JKGGpHc-Ham} denotes an external magnetic field which is used to tune magnetic phase transitions.
Conceptually, it is called [111] magnetic field if $\mathbf{h}$ is normal to the honeycomb plane.

The interest in the Kitaev materials can trace back to the synthesis of $5d^5$ iridates Na$_2$IrO$_3$ and $\alpha$-Li$_2$IrO$_3$
shortly after the proposal of Jackeli-Khaliullin mechanism \cite{Jackeli2009PRL}.
By using a topotactic replacement of interlayer alkali Li in the latter with hydrogen or silver ions,
alternative compositions of H$_3$LiIr$_2$O$_6$ and Ag$_3$LiIr$_2$O$_6$ are obtained \cite{Kitagawa2018Nature,Bahrami2019PRL}.
In parallel, $\alpha$-RuCl$_3$ with $4d^5$ configuration is appreciated as another remarkable Kitaev material
in that it has unusual broad low-energy magnetic excitations \cite{Banerjee2016NM}
and hosts field-induced intermediate disordered phase \cite{Sears2017PRB,Baek2017PRL,Wolter2017PRB,Wang2017PRL,Zheng2017PRL,Do2017NP,Ran2017PRL,WinterNcom2018,Ran2022CPL}.
Further, research efforts also identify the $3d^7$ materials like cobaltates Na$_2$Co$_2$TeO$_6$, Na$_3$Co$_2$SbO$_6$, and BaCo$_2$(AsO$_4$)$_2$
as Kitaev materials although they may have weaker spin-orbit coupling \cite{Songvilay2020PRB,Lin2021NC,Li2022PRX,Zhong2020SA}.
Subtly, the weak spin-orbit interaction is conducive to cause the formation of spin-orbit entangled $J_{\rm eff} = 1/2$ pseudospins
which are critical for producing Kitaev interactions.
Beyond these, Kitaev materials with higher spins have also been proposed,
including spin-1 $\rm{X_3Ni_2SbO_6 \left(X=Li, Na\right)}$ \cite{Zvereva2015PRB}, Na$_3$Ni$_2$BiO$_6$ \cite{Shangguan2023NP}, KNiAsO$_6$ \cite{Taddei2023PRR},
and spin-$3/2$ CrI$_3$ and $\rm{CrXTe_3 \left(X=Si,Ge\right)}$ \cite{Xu2018npjCM,Xu2020PRL,Stav2021PRR}.

In fact, value of the Kitaev interaction, together with other terms, is rather contrasting in different materials.
As such, study in this field has developed along with several particularly opportune aspects,
of which the prominent examples are the Kitaev-$\Gamma$ model and the $\Gamma$-$\Gamma'$ model.
We postpone the illustration of these models to subsequent Sections, and in what follows we will focus on the pure Kitaev and $\Gamma$ limits concisely.

\textit{2.1. Kitaev Honeycomb Model}.
The Kitaev honeycomb model consists of bond-directional Ising couplings between nearest-neighbor sites,
with the quantization axis depending on the bond type \cite{Kitaev2006AP}.
The Hamiltonian reads
\begin{eqnarray}\label{EQ:Kitaev-Ham}
\mathcal{H} = \sum_{\left<ij\right>\parallel{\gamma}} K_{\gamma} S_i^{\gamma}S_j^{\gamma},
\end{eqnarray}
in which $K_{\gamma}$ denotes the corresponding coupling constant along each bond type.
By rewriting spin operators via four Majorana fermions $\{b_i^x, b_i^y, b_i^z, c_i\}$ such that $S_i^{\gamma} = \imath b_i^{\gamma}c_i/2$,
the Hamiltonian in Eq.~\eqref{EQ:Kitaev-Ham} can be rewritten as
$\mathcal{H} = \frac{1}{4} \sum_{\left<ij\right>\parallel{\gamma}} b_i^{\gamma}b_j^{\gamma} c_ic_j$.
Since $b_i^{\gamma}$ fermions are local entities, the bond variable $u_{ij} = \imath b_i^{\gamma}b_j^{\gamma}$ is thus a constant of motion that takes $\pm1$,
a $\mathbb{Z}_2$ conserved quantity.
Therefore, the Hamiltonian can be further recast into a quadratic form
$\mathcal{H} = -\frac{\imath}{4} \sum_{\left<ij\right>} \langle u_{ij}\rangle c_ic_j$.
Clearly, for any given configuration of $\{u_{ij}\}$, the model can be diagonalized exactly.
In light of the Lieb's theorem, the ground state is found to be in the flux-free sector with hexagonal plaquette operator \cite{Kitaev2006AP}
\begin{eqnarray}\label{EQ:HexPlqttOp}
W_p = 2^6 S_1^xS_2^yS_3^z S_4^xS_5^yS_6^z
\end{eqnarray}
being $+1$ on every six-site hexagonal plaquette.
Since $W_p$ does not commute with local spin operators, the ground state cannot possess any long-range magnetic order.
This, in turn, suggests that the ground state of the Kitaev model is a kind of QSL with extremely short-range spin-spin correlations.
Depending on the relative intensity of $K_{\gamma}$,
the fermionic excitations could be either gapless (when $|K_x| \leq |K_y| + |K_z|$, $|K_y| \leq |K_z| + |K_x|$,
and $|K_z| \leq |K_x| + |K_y|$) or gapped otherwise \cite{Kitaev2006AP}.

The isotropic Kitaev model has received considerable attention, both from ground-state aspect and thermodynamic perspective.
Notwithstanding gapless fermionic excitations due to itinerant Majorana fermions,
the $\mathbb{Z}_2$ vison flux excitations are gapped and the latest exact calculation yields a gap energy of $\Delta_v = 0.263313(6) K$ \cite{Panigrahi2023PRB}.
Moreover, there are two pronounced peaks in the specific heat at well-separated temperature scales,
which can be understood from the two species of Majorana fermions \cite{Nasu2015PRB}.
The high-temperature crossover, driving by itinerant Majorana fermions, is intimately connected with the evolution of nearest-neighbor spin correlations.
By contrast, the low-temperature crossover stems from thermal fluctuations of fluxes comprised of localized Majorana fermions \cite{Nasu2015PRB}.

\textit{2.2. Honeycomb $\Gamma$ Model}.
The honeycomb $\Gamma$ model is a close cousin of the celebrated Kitaev honeycomb model shown in Eq.~\eqref{EQ:Kitaev-Ham}.
Due to the complexity of the interaction in the honeycomb $\Gamma$ model,
it is instrumental in firstly figuring out the underlying spin patterns of the ground state at the classical level.
By parameterizing each spin as $\mathbf{\hat S}_n = (\eta_i a, \eta_j b, \eta_k c)$ in which $\sqrt{a^2 + b^2 + c^2} = S$ and $\eta_{i,j,k}$ are independent Ising variables, Rousochatzakis and Perkins found that the ground-state degeneracy of a $N$-site cluster is $2^{N/2}$ in addition to free choices of $(a, b, c)$.
Therefore, the macroscopic degeneracy contributes to the formulation of classical spin liquid in its ground state \cite{Rousochatzakis2017PRL}.
On the other hand, the huge ground-state degeneracy of the antiferromagnetic (AFM) $\Gamma$ model is a source of unconventional phases,
which could be selected due to the interplay of exchange frustration and competing interactions.
Specifically, the 4-site zigzag phase displaying a peak at $\textbf{M}$ point
and the 18-site triple-meron crystal \cite{Chen2023NJP,Rayyan2021PRB} displaying a peak at $2\textbf{M}/3$ point
both belong to the degenerate manifold.
Methods to pick up several common phases from the degenerate manifold are shown in Table~\ref{Tab-DegGSL}.
We emphasize that when perturbing the $\Gamma$ limit by a ferromagnetic (FM) Kitaev interaction,
the resultant magnetic ordering is found to be the triple-meron crystal
by using the classical Monte Carlo method \cite{Chen2023NJP,Rayyan2021PRB} and the machine learning approach \cite{Liu2021PRR}.
We also note in passing that the anisotropy of the $\Gamma$ term among its three bonds cannot ruin the classical spin liquid,
as demonstrated by the Luttinger-Tisza analysis (not shown).

\begin{table}[th!]
\caption{Selected degenerate magnetic orderings out of the degenerate manifold of the classical honeycomb $\Gamma$ model.
  The number of sites in the unit cell (U.C.), the ordering wave vector $\textbf{Q}$, and the perturbation to pick up this phase are shown correspondingly.
  }\label{Tab-DegGSL}
  \begin{tabular}{ c c c c}
  \hline\hline\hline
  Phase                     & U.C.      & $\textbf{Q}$                              & Perturb.              \\
  \hline
  AFM phase                 & 2         & $\boldsymbol{\Gamma}'$                    & $J, \Gamma' > 0$      \\
  zigzag phase              & 4         & $\textbf{M}$                              & $J < 0$               \\
  120$^{\circ}$ phase       & 6         & $\textbf{K}$                              & $\Gamma' < 0$         \\
  counter-rotating spiral   & 6         & $\textbf{K}$ \& $\boldsymbol{\Gamma}'$    & $K > 0$               \\
  triple-meron crystal      & 18        & $2\textbf{M}/3$                           & $K < 0$               \\
  \hline\hline\hline
\end{tabular}
\end{table}

The Ising $\eta$ variables can be splitted into three interpenetrating triangular sublattices.
In the presence of thermal fluctuations, the emergent spherical symmetry at high enough temperature is lost,
compelling spins primarily point toward one of the six cubic directions.
After a suitable finite-size scaling of plaquette order parameter $\tilde{W}(\textbf{Q}) \propto \sum_{p} W_p e^{\imath\textbf{Q}\cdot\textbf{r}_{p}}$,
it is revealed that there is a hidden plaquette order in the classical spin-liquid ground state
and the temperature-driven thermal phase transition likely belongs the 2D three-state Potts universality class \cite{Saha2019PRL}.

However, the quantum ground state of the honeycomb $\Gamma$ model is rather involved,
as different methods obtain paradoxical results.
The exact diagonalization calculation on 24-site cluster and infinite density-matrix renormalization group (DMRG) study of a cylinder with a width of three unit cells
both declare that the ground state is a nonmagnetic phase \cite{Catuneanu2018npjQM,Gohlke2018PRB}.
Nevertheless, the infinite DMRG study also found a lattice-rotational symmetry breaking, leading to a nematic paramagnet \cite{Gohlke2020PRR}.
By contrast, the variational Monte Carlo simulation suggests that it is a zigzag phase \cite{Wang2019PRL}
while the pseudofermion functional renormalization group proposes that it is an incommensurate phase displaying peaks at nearly $2\textbf{M}/3$ point in its reciprocal space \cite{Buessen2021PRB}.
It is worth noting that the incommensurate phase is plausibly the triple-meron crystal
given that they share a similar landscape in the momentum space.
In some senses, in view of the tendency of each quantum many-body approach,
it may partially explain why different methods can target discrepant states.
As an attempt to resolve puzzle of the quantum ground state of the honeycomb $\Gamma$ model,
a large-scale DMRG calculation was performed on a wide cylinder with up to 200 sites
and the ground state is found to be a gapless $\Gamma$SL \cite{Luo2021NPJ}.
The details are reviewed in the subsequent Section.

\textit{3. Gapless $\Gamma$SL Revealed in Bond-modulated Heisenberg-$\Gamma$ model}.
To check for the tendency of zigzag ordering, Luo \textit{et al.} proposed a bond-modulated Heisenberg-$\Gamma$ model \cite{Luo2021NPJ} whose Hamiltonian reads
\begin{eqnarray}\label{EQ:BMJG-Ham}
\mathcal{H} = \tilde{J} \sum_{\left<ij\right>\parallel{\gamma}} \eta_{\gamma}\textbf{S}_i\cdot\textbf{S}_j
+ \Gamma \sum_{\left<ij\right>\parallel\gamma}(S_i^{\alpha}S_j^{\beta}+S_i^{\beta}S_j^{\alpha}),
\end{eqnarray}
in which $S_i^{\gamma}$~($\gamma$ = $x$, $y$, $z$) is the $\gamma$-component of the spin operator, and $\alpha$ and $\beta$ are the remaining two bonds.
In the conventional XC clusters, $\eta_{\gamma} = 1$ for the horizontal bonds and equals to $-1$ for the zigzag-directional bonds.
The exchange couplings $\tilde{J}$ and $\Gamma$ are parameterized as
$\tilde{J} = \cos\vartheta$ and $\Gamma = \sin\vartheta~(\geq 0)$ with $\vartheta\in[0,\pi]$.
As the angle $\vartheta$ varies, the classical ground state changes from zigzag phase to stripy phase,
passing through a mixed phase and a narrow noncollinear region.
By performing the DMRG calculation on four XC cylinders with the circumference $n$ ranging from 4 to 10,
it is found that the ground-state energy is somewhat flat in the middle region and displays sharp downwarping when away from the middle.
This indicates that there should be an intermediate region sandwiched between the zigzag phase and stripy phase.
Of note is that the ground-state energy per site $E_g$ of the honeycomb $\Gamma$ model is rather challenging to estimate,
in that it exhibits a nonmonotonic scaling behavior as the circumference $n$ increases, akin to that of the Kitaev honeycomb model.
However, the energy is linearly decreasing with length $L_x$ of the cylinder at each fixed circumference $n$.
After comparing the extrapolated energy at the achievable circumferences,
one can infer that the ground-state energy of the honeycomb $\Gamma$ model is $-0.354(3)$ in the thermodynamic limit.

\begin{figure}[!ht]
\centering
  \includegraphics[width=0.90\columnwidth, clip]{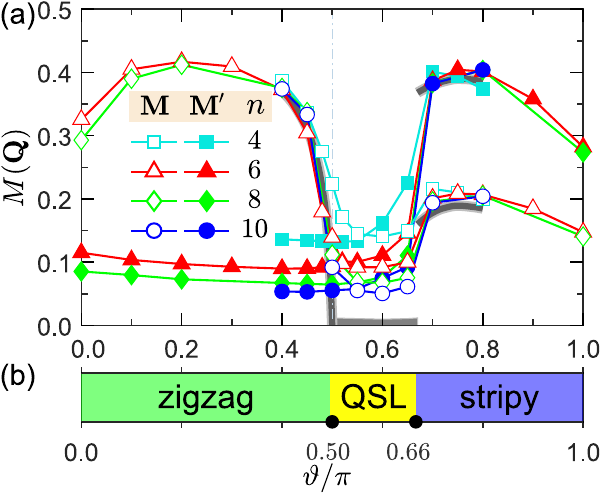}\\
  \caption{(a) Magnetic order parameters $M(\textbf{Q})$ for the zigzag order (open symbols) and stripy order (filled symbols)
    with $\textbf{Q} = \textrm{\bf{M}}$ and/or $\textrm{\bf{M}}'$ under four finite XC clusters.
    The thick gray line shows the magnetic order in the thermodynamic limit.
    (b) Quantum phase diagram of the bond-modulated $\tilde{J}$-$\Gamma$ model, which contains zigzag phase, QSL, and stripy phase \cite{Luo2021NPJ}.
    }\label{FIG01-BMJG}
\end{figure}

The magnetic spin structure factor (SSF) $\mathbb{S}_N({\bf{Q}})$, together with the associated magnetic order parameter  $M_N({\bf{Q}}) = \sqrt{\mathbb{S}_N({\bf{Q}})/N}$,
can be exploited to pin down the interval of the intervening phase.
The zigzag phase peaks at $\textbf{M}$ point while the stripy phase peaks at both $\textbf{M}$ and $\textbf{M}'$ points, see Fig.~\ref{FIG01-BMJG}(a).
By contrast, the intervening phase only displays a diffuse pattern with soft peaks in its magnetic SSF, signifying a QSL behavior.
After a careful inspection of the DMRG results on both cylinders and hexagonal clusters,
it is inferred that the intervening QSL phase persists in the window of $\vartheta_{t,l} \leq \vartheta \leq \vartheta_{t,r}$
with $\vartheta_{t,l} \simeq 0.50\pi$ and $\vartheta_{t,r} = 0.66(1)\pi$, see Fig.~\ref{FIG01-BMJG}(b).
Right at the $\Gamma$ limit where $\vartheta = \pi/2$, the magnetization decreases algebraically with the circumference $n$.
The linear extrapolation of the maximal peaks yields a vanishing magnetization, advocating the QSL phase therein.

The excitation gaps, $\Delta_{\upsilon}$ = $E_{\upsilon} - E_g$,
also exhibit distinguishing behaviors in magnetically ordered phases and the QSL phase.
On a 24-site hexagonal cluster, both zigzag and stripy phases own doubly degenerate ground states,
and thus $\Delta_1$ is vanishing small while $\Delta_2$ survives.
However, the ground state of the intermediate QSL phase is unique and density of state in the low-energy spectrum is denser when compared with its neighbor gapped phases.
Numerically, this collapse of excitation gaps can be interpreted as a signature of gapless spectrum.
The fact that lowest excitation gap at $\vartheta = \pi/2$ decreases from 0.08656249 ($N = 24$) to 0.04009844 ($N = 32$) is also consistent with a critical system.

The von Neumann entanglement entropy $\mathcal{S}_{\rm vN}$ is another versatile tool in diagnosing quantum criticality.
Depending on the width of the cylinder, the entanglement entropy displays a striking difference.
On a three-leg cylinder, $\mathcal{S}_{\rm vN}$ is exceedingly insensitive to the length, revealing a zero central charge.
By contrast, $\mathcal{S}_{\rm vN}$ increases with the length on a four-leg cylinder,
and the logarithmic fitting of $\mathcal{S}_{\rm vN}$ with respect to length $L_x$ gives a central charge of around 3.
Regarding the nature of QSL, the fact that the central charge depends highly on the width of cylinders may
either imply the existence of spinon Fermi surface whose pockets might be detected by different cuts in the Brillouin zone,
or hint the Dirac QSL with three Dirac Fermions around $\textbf{M}$ points.
Taken together, the vanishing magnetization, the collapsing excitation gaps, and the width-dependent central charge on long cylinders
manifest the ground state of honeycomb $\Gamma$ model is plausible a gapless QSL.

\textit{4. Spin-1 Kitaev-$\Gamma$ model}.
The Kitaev-$\Gamma$ model has captured massive attention both because of its relevance to real materials, and because of its ability to harbour exotic phases.
By exerting the perturbation theory on a five-orbital Hubbard model and exploiting the energy bands via the \textit{ab initio} calculation \cite{Wang2017PRB},
it is proposed that the effective Hamiltonian of $\alpha$-RuCl$_3$ is the Kitaev-$\Gamma$ model whose Hamiltonian is
\begin{eqnarray}\label{EQ:KGHam}
\mathcal{H} = \sum_{\left<ij\right>\parallel\gamma} \Big[K S_i^{\gamma} S_j^{\gamma}
    + \Gamma \big(S_i^{\alpha}S_j^{\beta}+S_i^{\beta}S_j^{\alpha}\big)\Big].
\end{eqnarray}
After fitting the inelastic neutron scattering data with the dynamical SSF,
it is estimated that the exchange couplings are $K = -6.8$ meV and $\Gamma = 9.5$ meV \cite{Ran2017PRL,Wang2017PRB}.
Although the convincing model to describe $\alpha$-RuCl$_3$  remains unclear \cite{Maksimov2020PRR},
it is widely believed that this model services as a reasonable starting point to understand the field-induced intermediate phase and nontrivial thermal Hall conductivity.
Theoretically, despite its deceptively simplicity,
the full phase diagram of the Kitaev-$\Gamma$ model with FM Kitaev interaction has not yet fully understood due to the strong frustration \cite{Catuneanu2018npjQM,Gohlke2018PRB,Gohlke2020PRR,Wang2019PRL,Buessen2021PRB}.
Starting from the FM Kitaev limit, it is still unclear which phase will appear next to the Kitaev QSL,
although some studies identify a ferromagnet due to the peak in the center of the first Brillouin zone.
For a review, see Ref.~\cite{Rousochatzakis2024RoPP}.
Pertinently, motivated by the synthesis of some higher-spin Kitaev magnets,
it is necessary to study the quantum phase transitions (QPTs) in the case of $S > 1/2$, say $S = 1$.

\textit{4.1. Chiral Spin State in Non-frustrated Region}.
Classically, the ground state of the non-frustrated region can be understood easily by a systematic analysis of spin patterns.
To commence with, it is beneficial to choose and parametrize a central spin as $\mathbf{S}_0 = (\eta_a a, \eta_b b, \eta_c c)$,
where ($\eta_a$, $a$) \big[respectively, ($\eta_b$, $b$) and ($\eta_c$, $c$)\big] denote the sign and intensity of $S_0^x$ (respectively, $S_0^y$ and $S_0^z$).
Here, $\sqrt{a^2 + b^2 + c^2} = S$ and $\eta$s are independent Ising variables that can be $\pm1$.
In view of peculiar structures of the Kitaev and $\Gamma$ interactions,
the three neighbor spins can be formulated as
$\mathbf{S}_i = -(\eta_a a, \eta_c c, \eta_b b)$ (for $\mathbf{X}$ bond),
$\mathbf{S}_j = -(\eta_c c, \eta_b b, \eta_a a)$ (for $\mathbf{Y}$ bond),
and $\mathbf{S}_k = -(\eta_b b, \eta_a a, \eta_c c)$ (for $\mathbf{Z}$ bond), respectively.
In this manner, entire spin textures can be generated accordingly,
giving rise to a (2+6)-ordering (the digits here represent the number of sites in unit cells)
which encompasses a two-sublattice AFM phase and a six-sublattice counter-rotating spiral order.
One should note that the AFM phase occurs only when $\eta_a = \eta_b = \eta_c = \pm1$ and $a = b = c = S/\sqrt3$.
Meanwhile, the spins in counter-rotating spiral order rotate clockwise on one of the two sublattices
while the remaining spins on the other sublattice rotate counter-clockwise around the [111] axis \cite{Stavropoulos2018PRB}.
Since there are merely three different $\eta$s, the ground-state degeneracy is eight at any fixed choice of $\{a, b, c\}$.
Despite the many possibilities for $\{\eta_a, \eta_b, \eta_c\}$ and $\{a, b, c\}$,
the classical ground-state energy per site is $e_{cl} = -(\Gamma + K/2)S^2$.

The counter-rotating spiral order can also be understood via the hidden SU(2) point.
To disclose it, it is essential to adopt a six-sublattice $\mathcal{T}_6$ transformation,
which is cyclic (anticyclic) permutations among the spin components at odd (even) sites within a unit cell \cite{Chaloupka2015PRB}.
In light of the $\mathcal{T}_6$ transformation, the Hamiltonian in Eq.~\eqref{EQ:KGHam} is rotated as
\begin{eqnarray}\label{EQ:KGHamT6}
\tilde{\mathcal{H}} = -\Gamma \sum_{\left<ij\right>} \mathrm{\tilde{S}}_i\cdot \mathrm{\tilde{S}}_j
                + (\Gamma-K)\sum_{\left<ij\right>\parallel\gamma} \tilde{S}_i^{\gamma} \tilde{S}_j^{\gamma},
\end{eqnarray}
in which the first term is the FM Heisenberg interaction while the second is the refined Kitaev interaction.
It should be emphasized that types of Ising bonds is redistributed after the rotation, reformulating a Kekul\'{e}-type pattern.
Specifically, $\mathbf{\tilde{X}}$ bonds orient outside, while $\mathbf{\tilde{Y}}$ and $\mathbf{\tilde{Z}}$ bonds alternate along the benzene-like ring.
When $K$ and $\Gamma$ are equal, Eq.~\eqref{EQ:KGHamT6} reduces to a SU(2)-symmetric FM Heisenberg model,
and the counter-rotating spiral order changes to be a dual FM phase after the rotation.

It is previsible that quantum fluctuations will further impose restrictions on the choice of $\{a, b, c\}$.
By using the real-space perturbation theory, the second-order calculation shows that the leading energy correlation is \cite{Rousochatzakis2024RoPP}
\begin{equation}\label{EQ:RSPTEg}
e_{cl}^{(2)} = -\frac{(\Gamma-K)^2S}{32|\Gamma+2K|}\left[\left(\frac{a}{S}\right)^4 + \left(\frac{b}{S}\right)^4 + \left(\frac{c}{S}\right)^4\right].
\end{equation}
Thus, when $K \neq \Gamma$, the total energy is minimized when one of the elements in $\{a, b, c\}$ is $S$ while the others are zero.
Also, only one of the $\eta$s that adheres to the nonzero element survives.
This result can be verified independently by using the linear spin-wave theory,
in which the spin-wave energy takes the minimal value on six points at the cubic axes.

\begin{figure}[!ht]
\centering
  \includegraphics[width=0.95\columnwidth, clip]{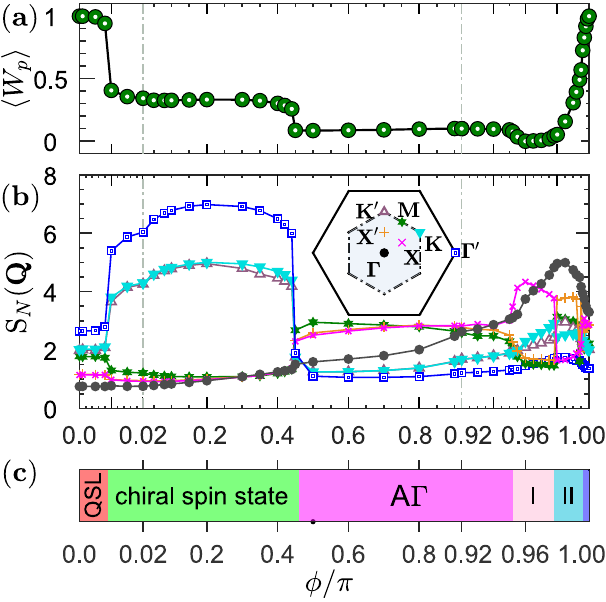}\\
  \caption{(a) The flux-like density $\langle\overline{W}_p\rangle$ as a function of $\phi$ on the 24-site hexagonal cluster.
  (b) The static SSF $\mathbb{S}_N({\bf{q}})$ at high-symmetry points shown in the reciprocal space (see inset).
  (c) The quantum phase diagram of the spin-1 Kitaev-$\Gamma$ model.
  The acronym A$\Gamma$ stands for an antiferromagnetic $\Gamma$ phase, and the symbols I and II represent two nematic ferromagnets \cite{Luo2024PRB}.
  Copyright 2024, American Physical Society.}\label{FIG-Spin1KG}
\end{figure}

At the quantum level, the emergent continuous symmetry is broken down to a discrete one,
and it has an interesting restriction on the magnetic order and scalar spin chirality.
It is straightforwardly to check that
$M(\boldsymbol{\Gamma'}) = S/2$ and $M(\textbf{K}) = S/\sqrt6$,
implying that $M(\textbf{K})/M(\boldsymbol{\Gamma'}) = \sqrt{6}/3 \approx 0.8165$.
As is shown in Fig.~\ref{FIG-Spin1KG}(b), such a relation has been verified by the DMRG calculation \cite{Luo2024PRB}.
Pertaining to the scalar spin chirality, it is defined as
\begin{equation}\label{EQ:ChiIJK}
\hat{\chi}^{\triangle}_{ijk} = \hat{\mathbf{S}}_i\cdot(\hat{\mathbf{S}}_j\times\hat{\mathbf{S}}_k),
\end{equation}
where sites ($i, j, k$) are three neighboring sites that belong to the same sublattice.
It follows that $\vert{\chi}_{ijk}\vert = S^3$ (which is a saturated value) since the three spins are perpendicular to each other.
It is in this sense that the underlying phase is dubbed a magnetically ordered chiral spin state.
In addition, the chiral spin state also owns a specific columnarlike or plaquettelike dimer covering, depending on the relative values of $K$ and $\Gamma$.
The hexagonal plaquette operator enjoys a trimerization phenomenon,
and the flux-like density $\langle \bar{W}_p \rangle = \langle \hat{W}_p \rangle/N_p$ ($N_p = N/2$)
is found to be $-1/3$ and $1/3$ for the spin-$1/2$ and spin-1 cases, respectively, see Fig.~\ref{FIG-Spin1KG}(a).

\textit{4.2. Nematic Ferromagnets in Frustrated Region}.
Determining the ground-state phase diagram near the FM Kitaev limit has been demonstrated to be an extremely challenging task.
By means of the von Neumann entropy and fluxlike density, it is observed that there are two distinct intervals in the neighbor of the Kitaev QSL.
In the SSFs, they have a dominating peak at $\boldsymbol{\Gamma}$ point and subleading peaks at $\textbf{K}/2$ and $\textbf{K}'/2$ points.
Since there are two different $\textbf{K}$ and $\textbf{K}'$ points in the Brillouin zone,
it is notable to emphasize that the specific positions of the subleading peaks are different in the two phases.
This discrepancy can not only be used to distinguish one to another, but also designate the likelihood of $C_3$ rotational symmetry breaking.
Coincidentally, there is an appreciable difference in the bond energy along the three different bonds for each phase,
showcasing a finite lattice nematicity defined as $\mathcal{O}_{\rm NF} = \vert\min(E^{\gamma}) - \max(E^{\gamma})\vert$.

The magnetic order parameter $M_N(\boldsymbol{\Gamma})$ and nematic order parameter $\mathcal{O}_{\rm NF}$ are amenable to pin down the character of phases.
On the one hand, by calculating $M_N(\boldsymbol{\Gamma})$ on four rhombic clusters with the total number of sites ranging from 12 to 32,
the magnetic order parameter at each exchange parameter is estimated to be a finite value in the thermodynamic limit.
On the other hand, while the extrapolation of $\mathcal{O}_{\rm NF}$ is infeasible,
it is found that $\mathcal{O}_{\rm NF}$ is quite robust against the geometry and site number of the cluster.
These evidences lead to the conclusion that these phases are nematic ferromagnets.
However, it is still unclear what is the major difference between the two, and this is an open question deserved to be studied in greater depth.

In the end, we wish to note that it is also interesting to study the parameter space in which $\Gamma < 0$.
Classically, by rotating $\mathbf{S}_i \rightarrow -\mathbf{S}_i$ on one sublattice,
we send $\theta \rightarrow \pi+\theta$ and thus one can only focus on $\Gamma > 0$.
However, this transformation is no longer applied in the quantum case.
Therefore, it is worthwhile to explore the potential novel phases at the quantum level in the case of negative $\Gamma$ term.

\textit{5. $\Gamma$-$\Gamma'$ Model: With and Without Magnetic Field}.
Since both $\Gamma$ and $\Gamma'$ interactions are off-diagonal exchange couplings,
the $\Gamma$-$\Gamma'$ model is thus expected to host exotic states of matter due to the interplay of exchange frustration and competing interactions.
Although this model is rather anisotropic, it can be rewritten as an easy-axis XXZ model with a $\mathbb{Z}_2 \ltimes U(1)$ symmetry when $\Gamma$ = $\Gamma'$,
indicating that the AFM phase and FM phase can be the ground state conditionally.
At the classical level, the entire phase diagram can be mapped out by the Lutinger-Tisza method.
There are four different phases, in which two of them are out-of-plane AFM phase (when $\Gamma' > 0$) and FM phase (when $\Gamma' < 0$)
while the other two are distinct in-plane 120$^{\circ}$ phases \cite{Luo2022PRB}.
Since the energy of the AFM phase is $-(\Gamma+2\Gamma')$ and the energy of the 120$^{\circ}$ phases is $-\left|\Gamma-\Gamma'\right|$,
it is inferred that there is a first-order phase transition between the 120$^{\circ}$ phase and the AFM phase
by switching the sign of the $\Gamma'$ interaction in the dominant $\Gamma$ region.
However, even within the framework of linear spin-wave theory,
the strong quantum fluctuations will alter the underlying phases and QPTs dramatically.
On the one hand, it is observed that the spin-wave energy of the zigzag phase is lower than that of the 120$^{\circ}$ phase,
implying that the zigzag phase may take the place of the 120$^{\circ}$ phase in the region of $\Gamma' < 0$.
On the other hand, energy reduction of the AFM phase is not satisfactorily large when compared with that of the zigzag phase in the vicinity of the $\Gamma$ limit.
Nonetheless, magnon instability of the latter will occur on conditions that $\Gamma'$ become positive,
hinting that there should be an intermediate region between the two magnetically ordered states \cite{Luo2022PRR}.

\textit{5.1 Chiral-spin Ordering}.
There are two exotic phases in the intermediate region,
in which one is the $\Gamma$SL protruding from the pure $\Gamma$ limit
while the other is a chiral-spin ordering with spontaneously broken time-reversal symmetry (TRS) \cite{Luo2022PRR}.
The chiral-spin ordering is characterized by a nonzero scalar spin chirality defined in Eq.~\eqref{EQ:ChiIJK}.
Depending on whether there is a lattice site in the center of the triangle, the chirality could be classified into two kinds,
i.e., $\chi_p$ (if it is centered at a site) and $\chi_n$ (if it is inside a hexagon).
For both kinds, the scalar spin chirality is uniformly distributed in the two sublattices but with a sign difference.
In addition, $\chi_p$ and $\chi_n$ have the same sign but different values within each sublattice.

\begin{figure}[!ht]
\centering
  \includegraphics[width=0.95\columnwidth, clip]{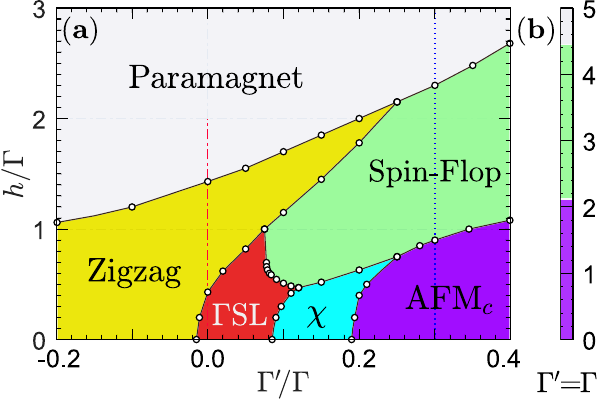}\\
  \caption{Quantum phase diagram of the $\Gamma$-$\Gamma'$ model in a [111] magnetic field.
  In the zero field, it contains zigzag phase, $\Gamma$SL, chiral-spin ordering (marked by the symbol $\chi$), and AFM phase.
  There are an intermediate spin-flop phase and a paramagnetic phase upon increasing the magnetic field \cite{Luo2022PRB}.
  Copyright 2022, American Physical Society.}\label{FIG-SpinFlop}
\end{figure}

On account of the TRS breaking, the ground state of the chiral-spin ordering is doubly degenerate
and its static SSF features a soft peak in the corner of the second Brillouin zone.
On top of the degenerate ground states, the lowest excitation gap changes from 0.11207 (24-site) to 0.00012 (32-site) on hexagonal clusters,
hinting a gapless excitation in the thermodynamic limit.
By calculating the entanglement entropy on a three-leg cylinder, the central charge is estimated as around 1.
Thus, the vanishing excitation gap and the finite central charge together support the existence of gapless excitations in the chiral-spin ordering.
In addition, as the system size increases, the magnetic order parameter vanishes after a proper extrapolation of the system size,
showing that the chiral-spin ordering is magnetically disordered.

The curious question is that whether the chiral-spin ordering is a QSL or not.
The dynamic SSF calculation finds a broad continuous feature in the low-frequency region, which is likely the evidence of the QSL phase.
However, the modular $\mathcal{S}$ matrix constructed from the minimal entangled states is trivial,
in violation of the topological order but is in line with a TRS-breaking chiral-spin ordering with short-range entanglement.
Hence, further studies are urgent to unveil the nature of the chiral-spin ordering.

\textit{5.2 Spin-flop phase}.
By switching on the [111] magnetic field, a spin-flop phase is identified to sandwich between
the low-field phases and the paramagnetic phase at large field, cf. Fig.~\ref{FIG-SpinFlop}.
Starting from the AFM phase, the spin-wave analysis predicts that the spin-flop phase is situated
in the field interval of $\sqrt{3}(\Gamma+2\Gamma') < h/S < 3(\Gamma+2\Gamma')$ \cite{Luo2022PRB}.
The spin flop refers to a first-order transition in which the ground state changes from the AFM ordering to the umbrella-like state
where the spins exhibit certain components that are normal to the field direction.
Noteworthily, the in-plane components are uniformly distributed, displaying a temporarily emergent $U(1)$ symmetry.
Nevertheless, the in-plane magnetization is unstable against extra perturbation,
and the continuous symmetry will break down to the discrete $C_3$ rotational symmetry,
accompanied by the occurrence of gapless Goldstone modes.
In fact, DMRG calculation suggests that excitations gaps are small and the spectrum is rather dense,
indicative of gapless excitations in the spin-flop phase.
When $\Gamma'$ = $\Gamma$, the $\Gamma$-$\Gamma'$ model in [111] magnetic field
is equivalent to an easy-axis XXZ antiferromagnet subjected to a longitudinal magnetic field.
By virtue of a proper mapping, it can be further reexpressed as an extended Bose-Hubbard model in the hard-core limit.
Therefore, the spin-flop phase is also recognized as a superfluid phase in the bosonic language.

Before closing, we note in passing that by exerting in-plane magnetic field in the $\Gamma$-$\Gamma'$ model,
two different spin-flop phases appear over the AFM and FM phases.
Depending on the relative value of the two exchange couplings as well as the strength and direction of the magnetic field,
it is demonstrated that within each spin-flop phase, there are at least three topologically distinct regions
in which the Chern number of the lowest magnon band belongs to set of $\{+1, 0, -1\}$ \cite{Li2024arXiv}.

\begin{figure*}[htb]
  \centering
  \includegraphics[width=0.85\linewidth, clip]{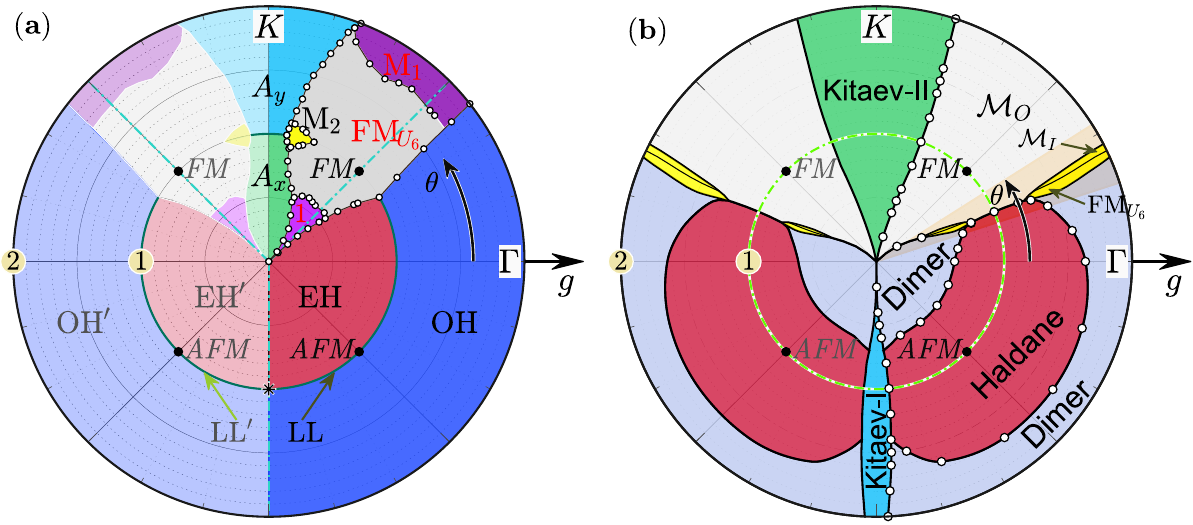}\\
  \caption{(a) Quantum phase diagram of the spin-$1/2$ Kitaev-$\Gamma$ chain.
  There are three magnetically ordered states, which are FM$_{U_6}$ phase, $M_1$ phase, and $M_2$ phase.
  There are also four magnetically disordered states, including $A_x$ and $A_y$ phases as well as even-Haldane (EH) and odd-Haldane (OH) phases.
  (a) Quantum phase diagram of the spin-1 Kitaev-$\Gamma$ chain.
  It contains FM$_{U_6}$ phase, $\mathcal{M}_I$ phase, $\mathcal{M}_O$ phase, dimerized phase, Haldane phase, and two distinct Kitaev phases \cite{Luo2021PRB,Luo2021PRR}.
  Copyright 2021, American Physical Society.}\label{FIG-ChainPD}
\end{figure*}

\textit{6. Anisotropic Kitaev-$\Gamma$ chain}.
Confronted with the difficulties of deriving convincing phase diagrams of the honeycomb-lattice Kitaev-$\Gamma$ model,
research effort has also been applied to 1D physics in attempts to capture referable hints.
The full Hamiltonian of the bond-alternating Kitaev-$\Gamma$ chain reads
\begin{equation}\label{BAKG-Ham}
\mathcal{H} = \sum_{l=1}^{L/2} g_x\mathcal{H}_{2l-1,2l}^{(x)}(\theta) + g_y\mathcal{H}_{2l,2l+1}^{(y)}(\theta)
\end{equation}
where the exchange term on each bond is given by
\begin{equation}\label{KtvGam-Ham}
\mathcal{H}_{i,j}^{(\gamma)}(\theta) = K S_i^{\gamma}S_j^{\gamma} + \Gamma (S_i^{\alpha}S_j^{\beta}+S_i^{\beta}S_j^{\alpha}).
\end{equation}
Here, $L$ is the length of the chain and $g_x$~($g_y$) denotes the strength of the odd~(even) bond.
By virtue of proper spin rotations, it is easy to verify that the model has two peculiar properties which strongly restrict the landscape of the phase diagram.
One is a self-duality relation which implies that each eigenvalue $E$ of $\mathcal{H}$ satisfies $E(g) = gE(1/g)$ with $g \equiv g_y/g_x$ being the relative bond strength,
while the other is a mirror relation that yields $E(K, \Gamma) = E(K, -\Gamma)$ \cite{Luo2021PRB}.

Further, a six-sublattice rotation is amenable to unveil a hidden SU(2) symmetry in the model when $K$ and $\Gamma$ are of the same strength.
Under this transformation the Hamiltonian casts into a diagonal form
\begin{equation}\label{U1XYZ-Ham}
\tilde{\mathcal{H}}_{i,j}^{(\gamma)}(\theta) = -K \tilde{S}_i^{\gamma} \tilde{S}_j^{\gamma}
- \Gamma (\tilde{S}_i^{\alpha} \tilde{S}_j^{\alpha}+ \tilde{S}_i^{\beta} \tilde{S}_j^{\beta})
\end{equation}
where the bonds $\gamma$ = $\tilde{x}$, $\tilde{z}$, and $\tilde{y}$ circularly \cite{Yang2020PRL}.
After parameterizing $\Gamma = \cos\theta$ and $K = \sin\theta$,
quantum phase diagrams of the spin-$1/2$ and the spin-1 Kitaev-$\Gamma$ are shown in Figs.~\ref{FIG-ChainPD}(a) and (b), respectively.

\textit{6.1 The $S = 1/2$ Case}.
The spin-$1/2$ bond-alternating Kitaev-$\Gamma$ chain has a rather rich phase diagram that hosts seven distinct phases.
These phases could be classified into three categories based upon their preferential interactions \cite{Luo2021PRB}.
Firstly, in the absence of the $\Gamma$ term, the model is reduced to the Kitaev spin chain
and is known to undergo a topological QPT between gapped $A_x$ and $A_y$ phases at $g = 1$.
On the AFM Kitaev side, regions of the $A_x$ and $A_y$ phases are enlarged by exerting $\Gamma$ interaction.
By contrast, these phases are dismissed immediately when the sign of the Kitaev interaction is negative.
Secondly, when $\theta = -\pi/4$ ($K = -|\Gamma|$), the model turns out to be a bond-alternating spin-$1/2$ Heisenberg chain which hosts two gapped disordered states
with different nonlocal string order parameters as long as $g \neq 1$.
The ground state is an odd-(even-)Haldane phase when $g > 1$~($g < 1$) and transition between the two is continuous with a central charge $c = 1$.
From a topological viewpoint, the odd-Haldane phase has fractional spin-1/2 edge states and is recognized as a symmetry-protected topological phase.

There are three magnetically ordered states which are in the neighbor of the hidden FM Heisenberg point where $K = |\Gamma|$.
In the vicinity of the isotropic limit, there is a FM$_{U_6}$ phase with an $O_h\to D_4$ symmetry breaking,
resulting in a six-fold degenerate ground states whose moments are along the $\pm\hat{x}$, $\pm\hat{y}$, and $\pm\hat{z}$ spin directions \cite{Yang2020PRL}.
For example, one of the moments within the unit cell could be
\begin{equation}\label{EQ:FMG0.80}
\big(\langle \tilde{\mathbf{S}}_1\rangle, \langle \tilde{\mathbf{S}}_2\rangle, \langle \tilde{\mathbf{S}}_3\rangle; \langle \tilde{\mathbf{S}}_4\rangle, \langle \tilde{\mathbf{S}}_5\rangle, \langle \tilde{\mathbf{S}}_6\rangle\big) \!=\! \big({c},{b},{b};{c},{a},{a}\big)\hat{z}
\end{equation}
where ${a}$, ${b}$ and ${c}$ are magnitudes of the spin orderings along the $\hat{z}$ direction and $a, b, c \leq S$.
When approaching the Kitaev side, there is a $M_2$ phase which has a spin-nematic correlation \cite{Luo2021PRB,Yang2021PRB}.
On the occurrence of large bond alternation, there is a magnetically ordered phase with an eight-fold ground-state degeneracy.
These degenerate states could be unified by the so-called $\eta$-notation \cite{Rousochatzakis2017PRL},
from which the spins within the six-site unit cell are
\begin{eqnarray}\label{EQ:M1Spin123}
\left(
    \begin{array}{c}
        \langle\tilde{\mathbf{S}}_1\rangle \\
        \langle\tilde{\mathbf{S}}_2\rangle \\
        \langle\tilde{\mathbf{S}}_3\rangle
    \end{array}
    \right)^T
    \!=\! \left[
    \left(
    \begin{array}{c}
        \eta_x a \\
        \eta_y b \\
        \eta_z c
    \end{array}
    \right),
    \left(
    \begin{array}{c}
        \eta_x a \\
        \eta_y c \\
        \eta_z b
    \end{array}
    \right),
    \left(
    \begin{array}{c}
        \eta_x c \\
        \eta_y a \\
        \eta_z b
    \end{array}
    \right)\right]
\end{eqnarray}
and
\begin{eqnarray}\label{EQ:M1Spin456}
\left(
    \begin{array}{c}
        \langle\tilde{\mathbf{S}}_4\rangle \\
        \langle\tilde{\mathbf{S}}_5\rangle \\
        \langle\tilde{\mathbf{S}}_6\rangle
    \end{array}
    \right)^T
    \!=\! \left[
    \left(
    \begin{array}{c}
        \eta_x b \\
        \eta_y a \\
        \eta_z c
    \end{array}
    \right),
    \left(
    \begin{array}{c}
        \eta_x b \\
        \eta_y c \\
        \eta_z a
    \end{array}
    \right),
    \left(
    \begin{array}{c}
        \eta_x c \\
        \eta_y b \\
        \eta_z a
    \end{array}
    \right)\right].
\end{eqnarray}
Here, $a$, $b$, $c$ $( \geq 0)$ satisfy the restriction $\sqrt{a^2+b^2+c^2} \leq S$,
while $\eta_{x}, \eta_y, \eta_z$ $( = \pm 1)$ are the Ising variables.
The three $\eta$'s are free to choose either 1 or $-1$ without altering the energy,
giving rise to the degenerate manifold.

\textit{6.2 The $S = 1$ Case}.
In similarity to the spin-$1/2$ case, the spin-1 bond-alternating Kitaev-$\Gamma$ chain also displays a phase diagram with seven distinct phases \cite{Luo2021PRR}.
In the Kitaev limit, there are two gapped Kitaev phases that both survive finite but different values of the $\Gamma$ interaction.
Consequently, region of the Kitaev phase near the FM Kitaev limit is much smaller than that of the AFM Kitaev side.
The ground states of the Kitaev phases are unique under periodic boundary condition while they are fourfold degenerate under open boundary condition,
analogous to that of the Haldane phase.
However, the fact that the lowest entanglement spectra in the Kitaev phases are unique is at odds with a symmetry-protected topological phase.
By increasing $\Gamma$ interaction from the FM Kitaev limit,
the Haldane phase appears when the bond alternation is modest and the dimerized phase is induced in the presence of large anisotropy.
Further, three distinct magnetically ordered states termed $M_O$, $M_I$, and FM$_{U_6}$ are favored when $\Gamma$ and Kitaev interactions are comparable and the latter is AFM.
The $M_O$ phase has an eightfold ground-state degeneracy and exhibits the same pattern shown in Eq.~\eqref{EQ:M1Spin123} and Eq.~\eqref{EQ:M1Spin456}.
In addition, extent of the $M_O$ phase is overwhelmingly larger than these of the $M_I$ and FM$_{U_6}$ phases.

While the spin-1 Kitaev chain is not exactly solvable, it is shown that it has an analog of the $\mathbb{Z}_2$-valued conserved quantities on bond.
The ground state locates at the flux-free sector where all the $\mathbb{Z}_2$ quantities are unitary,
while the degeneracy of the first excited state equals to the site numbers of the chain, corresponding to flipping the sign of arbitrary $\mathbb{Z}_2$ quantities.
Due to the unusual excitations, the Kitaev phase displays a nontrivial double-peak structure in the specific heat.
In contrast to the spin-$1/2$ Kitaev honeycomb model whose low-temperature peak comes from the freezing of vison fluxes,
the low-temperature peak in the spin-1 Kitaev chain relates to the freezing of $\mathbb{Z}_2$ quantities and is relevant to the highly degenerate low-lying excited states.
Moreover, the thermal entropy is released gradually without generating a plateau during the crossover temperature region,
which is again different from the Kitaev honeycomb model.

\textit{6.3 Insights of the Kitaev Phase}.
As can be seen from Fig.~\ref{FIG-ChainPD}(b), the bond alternation itself can not induce a QPT in the pure Kitaev limit.
However, the Kitaev phase will be taken over by the spontaneous symmetry breaking phase when the single-ion anisotropy (SIA) is brought in.
To interpolate between the [001]-type and [111]-type SIAs, one can perturb the Kitaev phase by the following term \cite{Luo2023PRB}
\begin{equation}\label{EQ:DzDcSIA}
\mathcal{H}_{\rm SIA} = D \sum_{i=1}^{L} \left[\frac{\sin\vartheta}{\sqrt2}(S_i^x+S_i^y)+\cos\vartheta S_i^z\right]^2
\end{equation}
with $D \leq 0$ and $\vartheta \in [0, \pi/2]$.
The full Hamiltonian respects a TRS, a dihedral symmetry, a translational symmetry, and so forth,
enabling the occurrence of certain spontaneous symmetry breaking phases.
In the Kitaev chain with a [001]-type SIA (i.e., $\vartheta = 0$) or a [111]-type SIA (i.e., $\vartheta = \tan^{-1}(\sqrt2)$),
the ground state undergoes a QPT to the dimerized phase or the AFM phase at $D = -0.6551(2)$ or $D = -0.6035(2)$, respectively.
Both QPTs belong to the Ising universality class in which the central charge $c = 1/2$, and the critical exponents $\beta = 1/8$ and $\nu = 1$.
The dimerized phase breaks the translational symmetry, while the AFM phase breaks the TRS and dihedral symmetry.
Thus, they can only have a first-order QPT in the Landau's paradigm.
However, for large enough SIA,
the dimerized phase and the AFM phase is found to undergo a ``Landau-forbidden" continuous QPT with a central charge $c = 1$.
This indicates that the spin-1 Kitaev chain can be a playground to support deconfined quantum criticality.

As a matter of fact, depending on the intensity and angle of the SIA, the Kitaev phase can be classified into two types.
Whereas it is characterized by a unique ground state under periodic boundary condition,
degeneracy of the first excited state could be either unique or doubly degenerate.
The first excited ground state is unique in the wide area (for example, in the region around the path of [111]-type SIA),
as compared to the twofold situation observed in a particular area where $|D|$ and $\vartheta$ are small.
It is in this sense that the Kitaev phase can be distinguished into type-I and type-II, respectively,
on account of the degeneracy of the first excited state.
Since the lowest excitation gap of the Kitaev phase never close throughout its entire area,
the change between the two types is regarded as a crossover rather than a QPT.
The two types can also be discriminated by the spatial distribution of bond-parity operator $\hat{W}_l$ in their energy spectra.
although $\hat{W}_l$ is totally flat in the ground state for the absent of translational symmetry breaking,
it is smoothly varied in type-I but undergoes a rapid oscillation in type-II in the excited states.

In spite of the differentiation in the Kitaev phase, the hallmark of its ground state is that it possesses a spin-nematic order parameter $\mathcal{O}_{SN}$,
which can be extracted from the four-spin correlation function
\begin{equation}\label{EQ:SNOP}
Q_{1}\big(i, j\big) = \langle S_{i}^{+} S_{i+1}^{+} S_{j}^{-} S_{j+1}^{-}\rangle \simeq \mathcal{O}_{SN}^2 e^{-\imath \phi},
\end{equation}
where $\phi$ is a phase factor that varies as the interaction strength changes.
Since the Kitaev phase preserves the translational symmetry and TRS,
$Q_{1}\big(r\big) \equiv Q_{1}\big(i, j\big)$ with $r = |j-i|$.
It is found that $Q_1(r)$ has a strong even-odd effect,
in which $\phi$ is finite when $r$ is odd and vanishes otherwise.
In a specific cut along the line of [111]-type SIA,
The nontrivial $\phi$ changes from $\pi$ to zero, as $|D|$ increases from the pure Kitaev limit.
Noteworthily, it experiences a rapid change near the quantum critical region,
The derivative of $\phi$ with respect to $D$, $\partial\phi/\partial D$,
is found to diverge at the quantum critical point as the chain length goes to infinity.
This indicates that the phase factor $\phi$ relating to $Q_{1}\big(r\big)$ ($r$ is odd) can serve as a tool to probe the QPT.

\textit{7. Summary and Perspective.}
We have reviewed some numerical works related to exotic phases and quantum phase diagrams in Kitaev-$\Gamma$ model and $\Gamma$-$\Gamma'$ model,
both of which are particular cases of the generic $JK\Gamma\Gamma'$ model.
In contrast to the long-standing SU(2)-symmetric models,
these models only have discrete symmetries, rendering it is extremely difficult to handle enough system size and achieve satisfactory numerical precision simultaneously.
Despite the challenges, people can still excavate many interesting phases including QSLs and others.
Although these studies do not necessarily resolve all the puzzles,
they at least demonstrate that the spin-orbit coupled model offers a fertile playground for exploring unconventional phases of matter
and comprehending collective emergent quantum phenomena.

Beyond the scope of current overview, there are several tantalizing and challenging issues remaining to be addressed further.
Firstly, in the spin-$1/2$ AFM Kitaev model on the honeycomb lattice,
the fate of the Kitaev QSL in presence of an [111] magnetic field is a topic of great interest,
and the number of and the nature of the intermediate phase before entering into the trivial polarized phase are under hot debates \cite{Zhu2018PRB,Hickey2019NC,Patela2019PNAS,Jiang2020PRL,Zhang2022NC,Wang2024arXiv}.
Secondly, while the ground state of the spin-1 Kitaev honeycomb model is widely believed to be a QSL,
whether its excitations are gapped or gapless remain unsettled \cite{Koga2018JPSJ,Oitmaa2018PRB,Lee2020PRR,Chen2022PRB}.
Thirdly, akin to the spin-$1/2$ analog,
it is interesting as well to pin down the nature of field-induced intermediate phase in the spin-$1$ AFM Kitaev model \cite{Zhu2020PRR,Hickey2020PRR,Khait2021PRR}.
Finally, it is also imperative to propose a Hamiltonian
that undergoes an in-plane field-induced QSL phase on top of a zigzag ordering in the multi-dimensional $JK\Gamma\Gamma'$ parameter space.

\textit{Acknowledgements.}
We thank Ken Chen, Jacob S. Gordon, Shijie Hu, Jinbin Li, Xudong Li, Ahmed Rayyan, P. Peter Stavropoulos, Bin Xi,
and Zongsheng Zhou for the collaboration on related topics, in particular for Hae-Young Kee.
This work is supported by the National Program on Key Research Project (Grant No. MOST2022YFA1402700),
the National Natural Science Foundation of China (Grants No. 12304176, No. 12274187, No. 12247183, and No. 12247101),
and the Natural Science Foundation of Jiangsu Province (Grant No. BK20220876).
The computations are partially supported by High Performance Computing Platform of Nanjing University of Aeronautics and Astronautics.

\end{document}